\documentclass[aps,prx,twocolumn,floatfix,longbibliography,nofootinbib,superscriptaddress]{revtex4-2}
\usepackage[utf8]{inputenc}
\usepackage{natbib}
\usepackage{graphicx}
\usepackage{tikz}
\usepackage{xcolor}
\usepackage[tbtags]{amsmath}

\definecolor{brightmaroon}{rgb}{0.76, 0.13, 0.28}

\usepackage[colorlinks, linkcolor=blue]{hyperref}
\hypersetup{colorlinks,allcolors=brightmaroon}
\usepackage{amssymb}
\usepackage{booktabs}
\usepackage{amsmath}
\usepackage{gensymb}
\usepackage{float}
\usepackage{amsmath}
\usepackage{tabularx,graphicx}
\usepackage{epstopdf}
\usepackage{latexsym}
\usepackage{color, colortbl}
\usepackage{psfrag}
\usepackage{bbm}
\usepackage{bm}
\usepackage{titlesec}
\usepackage{dsfont}
\usepackage{feynmp}
\usepackage{slashed}
\usepackage{multirow}
\usepackage[normalem]{ulem}
\renewcommand{\vec}[1]{\boldsymbol{#1}}

\newcommand{\bcen}{\begin{center}}
\newcommand{\ecen}{\end{center}}
\newcommand{\btab}{\begin{tabular}}
\newcommand{\etab}{\end{tabular}}
\newcommand{\bdes}{\begin{description}}
\newcommand{\edes}{\end{description}}

\newcommand{\beq}{\begin{equation}}
\newcommand{\eeq}{\end{equation}}
\newcommand{\bea}{\begin{eqnarray}}
\newcommand{\eea}{\end{eqnarray}}

\newcommand{\half}{\frac{1}{2}}
\newcommand{\bary}{\begin{array}}
\newcommand{\eary}{\end{array}}
\newcommand{\benum}{\begin{enumerate}}
\newcommand{\eenum}{\end{enumerate}}
\newcommand{\bitem}{\begin{itemize}}
\newcommand{\eitem}{\end{itemize}}

\newcommand{\bfig}{\begin{figure}}
\newcommand{\efig}{\end{figure}}
\renewcommand{\vec}[1]{\boldsymbol{#1}}
\newcommand{\al}{\alpha}

\newcommand{\eqn}[1] {Eq.~(\ref{#1})}

\newcommand{\Fig}[1]{Fig.~\ref{#1}}

\makeatletter

\newcommand{\Rmnum}[1]{\expandafter\@slowromancap\romannumeral #1@}
\makeatother

\begin{document}

\title{Diffusive metal in a percolating Chern insulator}


\author{Subrata Pachhal}
\email{pachhal@iitk.ac.in}
\affiliation{Department of Physics, Indian Institute of Technology Kanpur, Kalyanpur, UP 208016, India}

\author{Naba P. Nayak}
\email{nabaprakash@iitb.ac.in}
\affiliation{Institut Langevin, ESPCI Paris, Université PSL, CNRS, 75005 Paris, France}
\affiliation{Department of Physics, Indian Institute of Technology Bombay, Powai, Mumbai 400076, India}

\author{Soumya Bera}
\email{soumya.bera@iitb.ac.in}
\affiliation{Department of Physics, Indian Institute of Technology Bombay, Powai, Mumbai 400076, India}

\author{Adhip Agarwala}
\email{adhip@iitk.ac.in}
\affiliation{Department of Physics, Indian Institute of Technology Kanpur, Kalyanpur, UP 208016, India}

\begin{abstract}
Two-dimensional non-interacting fermions without any anti-unitary symmetries generically get Anderson localized in the presence of disorder. In contrast, topological superconductors with their inherent particle-hole symmetry can host a thermal metallic phase, which is non-universal and depends on the nature of microscopic disorder. In this work, we demonstrate that in the presence of geometric disorders, such as random bond dilution, a robust metal can emerge in a Chern insulator with particle-hole symmetry. The metallic phase is realized when the broken links are weakly {\it stitched} via concomitant insertion of $\pi$ fluxes in the plaquettes. These nucleate low-energy manifolds, which can provide percolating conduction pathways for fermions to elude localization. This diffusive metal, unlike those in superconductors, can carry charge current and even anomalous Hall current. We investigate the transport properties and show that while the topological insulator to Anderson insulator transition exhibits the expected Dirac universality, the metal insulator transition displays a different critical exponent $\nu \approx 2$ compared to a disordered topological superconductor, where $\nu \approx 1.4$. Our work emphasizes the unique role of geometric disorder in engineering novel phases and their transitions in topological quantum matter.
\end{abstract}

\maketitle

{\it Introduction.---}Fundamental understanding of the symmetry-protected topological (SPT) phases is rooted in crystalline band theory and their tenfold classification based on underlying symmetries \cite{Hasan_RMP_2010, Qi_RMP_2011, Shen_Book_2013, Bernevig_Book_2013, Altland_PRB_1997, Kitaev_AIP_2009, Ludwig_PS_2015}. In two dimensions, Chern insulators are one such example characterized by a non-trivial Chern number $\mathcal{C}$ of the bulk band and a corresponding quantized anomalous Hall response $\sigma_{xy} = \frac{e^2}{h}\mathcal{C}$ resulting from broken time-reversal symmetry \cite{He_QAH_2013, Liu_QAH_2016}. With the ever-expanding number of topological materials \cite{bernevig_nature_2017, bernevig_topmat_database}, interest in the effects of disorder has surged in recent years both theoretically \cite{Li_TAI_2009, Groth_PRL_2009, Prodan_PRL_2010, Kobayashi_PRL_2013, Agarwala_PRL_2017, Prodan_2017,  Li_disorderedQAH_2021, Atland_AOP_2023, Mondal_PRB_2023, mondal_IRFP_2025} and experimentally \cite{Paul_amorphusbi2se3_2023, Liu_PRL_2017, stutzer2018photonic, Liu_TAI_photonic_2020, LiPRL2021, costa_amorphusti_2019, zhou_amorphusTI_2020, Zhang_amorphustopology_science_2023}. Remarkably, in class D topological superconductors (TSC), which possess only particle-hole symmetry, field-theoretic analysis predicts that the disorders generically drive the topologically insulating phase into a heat-conducting {\it thermal} metal defying Anderson localization \cite{Senthil_PRB_2000}. In contrast, in the context of the random-bond Ising model \cite{Cho_PRB_1997}, which has a class D fermion representation, network models show that the emergence of the metallic phase strongly depends on the very nature of the disorder itself \cite{Read_RBIM_nometal_2000, Chalker_PRB_2001}. While a spatially fluctuating energy scale, such as in Anderson disorder, supports the thermal metal phase \cite{Midenberger_PRB_2007, Medvedyeva_PRB_2010, Laumann_PRB_2012, Fulga_andersonbond_PRL_2020, Wang_PRB_2021}; disorder generated via geometric origins, such as via percolation \cite{Aharony_T&F_2003} or structural defects, shows localization \cite{Ivaki_PRR_2020, Sahlberg_PRR_2020}. Thus, the origin of a metallic phase in class D SPTs, and the interplay of `geometric' disorder and `energetic' disorder in generating them, has been far from clear.

\begin{figure}
    \centering
    \includegraphics[width=0.85\linewidth]{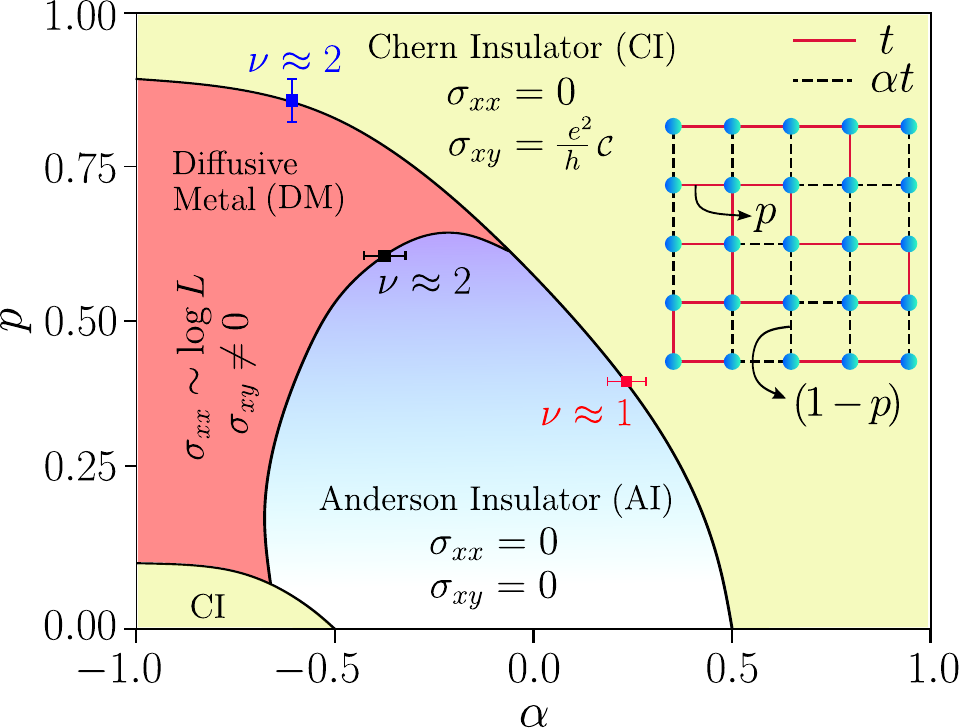}
    \caption{\textbf{Phase diagram:} Schematic phase diagram of a random bond-disordered Chern insulator. The inset shows the disorder protocol where the hopping strengths are either $t$ with probability $p$ or $\alpha t$ with probability $(1-p)$. Apart from Chern insulator (CI) and localized Anderson insulator (AI), the phase diagram features a diffusive metal (DM) phase characterized by longitudinal conductivity $\sigma_{xx} \sim \log L$ ($L$ is the system size), and non-quantized $\sigma_{xy}$. While for DM-AI and DM-CI transition critical exponent $\nu \approx 2$, for AI-CI transition $\nu \approx 1.$}
    \label{fig1}
\end{figure}

In this work, we pose---in a geometrically disordered class D Chern insulator, what decides the Anderson localization? Can disorder drive these localized electrons into a metallic phase? To address these questions, we consider a microscopic model of a class D insulator with $\mathcal{C} = 1$, defined on a square lattice that is punctured via random bond dilution. We then introduce a {\it weak} coupling $\alpha$ to {\it stitch} the {\it broken} bonds such that each bond has strength $t$ probability $p$ and $\alpha t$ with probability $(1-p)$ as shown in the inset of \Fig{fig1} and dub the system as ``random bond-disordered Chern insulator'' (RBCI). With extensive numerical analysis of quantum transport properties such as longitudinal conductivity $\sigma_{xx}$ and Hall response $\sigma_{xy}$, we uncover the rich phase diagram of the RBCI in the $\alpha$-$p$ space, as shown schematically in \Fig{fig1}. In the $\al \in [0, 1]$ regime, the phase diagram exhibits a localized Anderson insulator (AI) phase along with the Chern insulator (CI) phase, consistent with earlier studies on geometrically disordered class D systems \cite{Ivaki_PRR_2020, Sahlberg_PRR_2020}. But remarkably, as we percolate the disconnected regions via {\it negative} $\alpha$ coupling, a diffusive metal (DM) phase emerges featuring logarithmic growth of $\sigma_{xx}$ with the linear size of the system. This diffusive metal arising in our RBCI system is electronic in nature and fundamentally differs from a thermal metal, where the quasiparticles are Majorana fermions and can carry only {\it heat}. Interestingly, unlike thermal metal \cite{Laumann_PRB_2012, Fulga_andersonbond_PRL_2020}, the DM phase supports an anomalous Hall response, $\sigma_{xy}$, attributed to broken time-reversal invariance.

Further employing finite-size scaling of the $\sigma_{xx}$ and $\sigma_{xy}$, we probe different phase transitions and their corresponding critical exponents. While AI to CI insulator-insulator transitions exhibit $\nu \approx 1$ localization length exponent, similar to the Dirac universality of the clean Chern transition, we reveal a distinct universality of the metal-insulator transitions with $\nu \approx 2$ for both DM-AI and DM-CI, in contrast to the $\nu \approx 1.4$ for a similar thermal metal-insulator transition in class D TSCs \cite{Victor_classDexponent_2008, Wang_PRB_2021}. We further explain that, in contrast to $\alpha > 0$, a negative $\alpha$ creates randomly distributed $\mathbb{Z}_2$ flux plaquettes in the lattice, which gives rise to zero-energy modes. At sufficiently high density, these sub-gap modes percolate throughout the system, resulting in global conducting channels for the localized electron and consequently rendering metallicity in the RBCI.

{\it Random bond-disordered Chern insulator.---}We consider spinless fermions, hopping between nearest-neighbor sites on a square lattice, governed by,
\begin{equation}
	H = \sum_{i, \vec{\eta}}\Big( \Psi^\dagger_{i} t_{i, \vec{\eta}} T_{\vec{\eta}}\Psi_{i+\vec{\eta}} + \text{h.c.}  \Big)+ \sum_i \Psi^\dagger_i \sigma_z \Psi_i,
	\label{eq_rham}
\end{equation}
where $\Psi_i = (c_{iA}~~ c_{iB} )^T$ with $c_{iA}$ and $c_{iB}$ representing fermionic annihilation operators for $A$ and $B$ orbital degree of freedom at $i^{\text{th}}$ position and $\vec{\eta}=(\hat{x}$, $\hat{y})$ denotes the unit vectors along $x$ and $y$ directions. Corresponding $\Psi^{\dagger}_i$ represents fermionic creation. Associated hopping matrices are $T_{\hat{x}}=-\half \left(\sigma_{z}+i\sigma_{x}\right)$ and $T_{\hat{y}}=-\half\left(\sigma_{z}+i\sigma_{y}\right)$, while
$\sigma_{z}$ accounts for the onsite orbital energy splitting, with $\sigma_{x/y/z}$ representing the Pauli matrices. The hopping strength for a given bond $t_{i, \vec{\eta}}$ is randomly chosen from the following bimodal distribution, 
\begin{equation}
    \mathcal{P} (t_{i, \vec{\eta}}) =  p\delta(t_{i, \vec{\eta}} - t) + (1-p)\delta(t_{i, \vec{\eta}} - \alpha t), \label{eq_bonddis}
\end{equation}
at a given probability $p$ and bond-factor $\alpha \in \left[-1, 1\right]$. While the system breaks time-reversal symmetry, it follows $H\rightarrow H$ under particle-hole operation, $\Psi_i \rightarrow \sigma_x[\Psi_i^\dagger]^T, ~  \Psi^\dagger_i \rightarrow [\Psi_i]^T \sigma_x$, restricting it to symmetry class D of {\it tenfold} classification \cite{Altland_PRB_1997}. 

Setting $t=1$, when all the bond strengths are uniform in the lattice, that is, at $p=1$ or $\alpha = 1$, the Hamiltonian in \eqn{eq_rham} can be written in momentum space as, $\mathcal{H}(\vec{k}) = \sin k_x \sigma_x + \sin k_y \sigma_y +  \left(1-\cos k_x-\cos k_y\right)\sigma_z$. This class D insulating system has non-trivial topology with Chern number $\mathcal{C} = 1$ \cite{BHZ_Sci_2006, QWZ_spinlessBHZ_2006}. At another clean limit $p = 0$, the system undergoes a topological phase transition via Dirac gap closing at $\alpha= \pm 0.5$ separating two topologically distinct phases: $\mathcal{C} = 0$ for $|\alpha| < 0.5$ and $\mathcal{C} = 1$ for $|\alpha| > 0.5$ (for details see Supplemental Material (SM) \cite{sm}).

Away from the clean limits (i.e., $0<p<1$) and at $\alpha = 0$, the Hamiltonian in \eqn{eq_rham} corresponds to bond-dilution disorder in a Chern insulator. In such a geometrically disordered system, a finite $\alpha$ essentially {\it stitches} the broken bonds as shown schematically in the inset of \Fig{fig1}, leading to a random-bond disordered Chern insulator (RBCI). Throughout our study, we set $t=1$ and consider half-filling of electrons such that the Fermi energy is pinned at {\it zero}.

\begin{figure}
    \centering
    \includegraphics[width=1\linewidth]{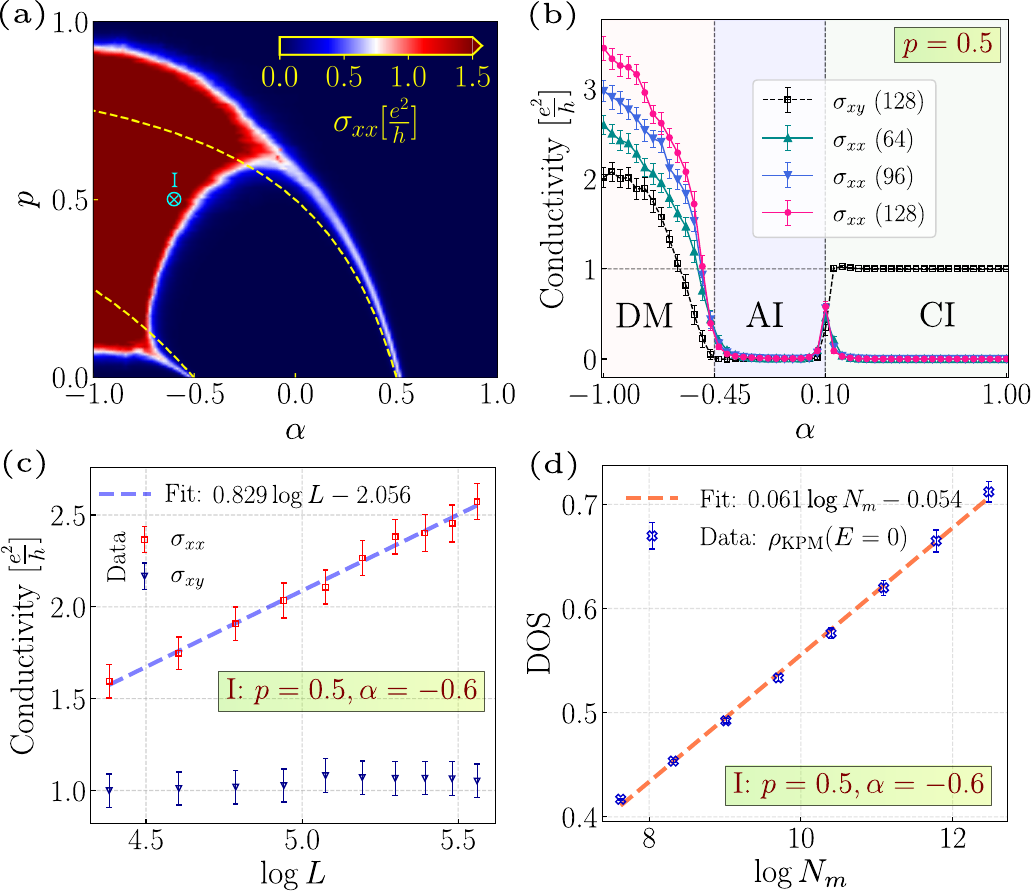}
    \caption{\textbf{Characterizing different phases:} (a) Configuration averaged $\sigma_{xx}$ phase diagram of the RBCI on a $40 \times 40$ lattice, computed over $200$ disorder realizations. Dashed lines denote the critical boundaries considering the virtual crystal approximation. (b) Along $p=0.5$, $\sigma_{xx}$, and $\sigma_{xy}$ distinguish DM, AI, and CI phases across $\alpha$. (c) The DM phase (point I: $p=0.5, \alpha = -0.6$) shows $\sigma_{xx}\sim \log L$ and finite $\sigma_{xy}$. Conductivities in (b, c) are averaged over 400 configurations; while for $\sigma_{xx}$ computation, width $W$ = $2L$, for $\sigma_{xy}$, $W = L/2$. (d) The disorder-averaged ($10^3$ configurations) density of states (DOS) for $L=512$ in the DM phase (point I) exhibits a logarithmic divergence.}
    \label{fig2}
\end{figure}

{\it Phases and their features.---}In the $\alpha$-$p$ parameter space, to obtain the phase diagram of the RBCI system, we evaluate both longitudinal and transverse conductivity $\sigma_{xx}$ and $\sigma_{xy}$ numerically using \texttt{KWANT} \cite{Groth_Kwant_2014}. While in a two-terminal rectangular geometry with periodic boundary conditions (PBC) in the transverse direction, lead-to-lead transmission at zero-energy divided by the aspect ratio gives $\sigma_{xx}$ in the unit of $e^2/h$, for $\sigma_{xy}$, we use a six-terminal device (see SM \cite{sm}). 

The configuration averaged $\sigma_{xx}$ in the $\alpha$-$p$ phase space is illustrated in \Fig{fig2}(a). The phase boundaries in the low $p$ limit can be captured via the {\it virtual crystal approximation} \cite{Soven_CPA_PR1967}, where the RBCI (\eqn{eq_rham}) is replaced by a translationally invariant Hamiltonian with average hopping of strength $(1-p)\alpha +p$ (see SM \cite{sm}). Following this, we obtain the {\it dashed} critical lines shown in \Fig{fig2}(a). Thus, two CI phases ($\mathcal{C}=1$) for $|\alpha| > 0.5$ at $p=0$ continue under small geometric fluctuations, while the intermediate trivial insulator ($|\alpha| < 0.5$) becomes a localized AI phase due to randomness in the hopping strength. At an intermediate $p$, while in the positive $\alpha$ regime the localized phase enters a CI phase, it surprisingly exhibits delocalization as $\alpha$ reduces below a certain negative value, giving rise to a metallic region (DM phase) in the phase diagram characterized by a macroscopic $\sigma_{xx}$ (see \Fig{fig2}(a)). In the $p\rightarrow1$ limit, the metallic regain again becomes insulating, and only the CI phase remains for all values of $\alpha$. The cut plot along $p=0.5$ shown in \Fig{fig2}(b) demonstrates these three phases distinctively: (1) Towards $\alpha \gtrapprox 0.1$, the insulating phase with $\sigma_{xx} \rightarrow 0$ and quantized Hall response, $\sigma_{xy} = e^2/h$ marked the CI phase. (2) In the regime $-0.45 < \alpha < 0.1$, both $\sigma_{xx}$ and $\sigma_{xy}$ are exponentially small indicating the AI phase. (3) Below a certain negative $\alpha$ ($\lessapprox -0.45$), $\sigma_{xx}$ increases with system size $L$ illustrating the diffusive metallic (DM) nature . Interestingly, in the DM phase, we find that not only $\sigma_{xx}$ follows $\log L$ behavior as shown in \Fig{fig2}(c) for a phase-space point $(p, \alpha)\equiv (0.5,-0.6)$, it also exhibits non-quantized Hall response $\sigma_{xy}$. 

Moreover, we employ the Kernel Polynomial Method (KPM) \cite{KPM_RMP_2006} to calculate the density of states (DOS) in our system. While in the metallic phase the disorder-averaged DOS at zero-energy $\rho_{\text{KPM}}(E=0)$ exhibits logarithmic divergence with number of moments $N_m$ in the Chebyshev expansion (see \Fig{fig2}(d)) which imply $\rho(E) \sim \log(1/E)$ as predicted for class D disordered systems \cite{Senthil_PRB_2000, Midenberger_PRB_2007}; in the AI phase, DOS displays a small spectral gap with localized states in the full spectrum (see SM \cite{sm}). Furthermore, we find that the metallic eigenstates are multifractal in nature (see SM \cite{sm}), which is generic in a two-dimensional disordered metal \cite{Falko_EPL_1995}. While these phases exhibit distinct interesting features, we now explore the nature of transitions between them.

{\it Phase transition and criticalities.---}A significant amount of scientific effort has been devoted over the years to understand the nature of transitions out of topological phase driven by disorder \cite{Victor_classDexponent_2008, Yamakage_PRB_2013, Wang_PRB_2021, Naba_QHC_2024}. With its two topologically distinct insulating phases and an intermediate metallic phase, the RBCI (see \Fig{fig1}(b)) offers a natural setting for investigating a variety of metal-insulator and insulator-insulator transitions in two dimensions.

\begin{figure}
    \centering
    \includegraphics[width=1\linewidth]{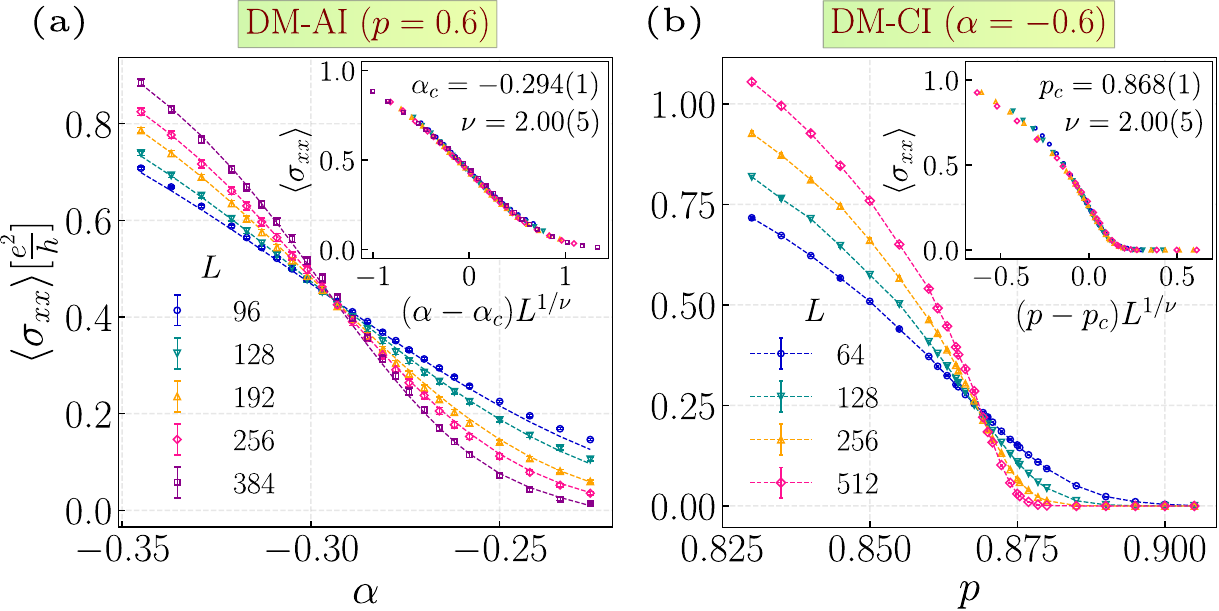}
    \caption{\textbf{Probing phase transitions:} (a) Finite size scaling of average $\sigma_{xx}$ across the DM-AI transition at $p=0.6$. Inset: Scaling collapse with $\alpha_c \approx -0.294$ and critical exponent $\nu \approx 2$. (b) Same as (a), but across the DM-CI transition with $p$ at $\alpha = -0.6$. Inset: scaling collapse with $p_c \approx 0.868$ and critical exponent $\nu \approx 2$. The standard error bars are within the width of the markers. See SM \cite{sm} for the number of configurations considered to compute disorder averaging.}
    \label{fig3}
\end{figure}

We first investigate the metal-insulator transition in the negative $\alpha$ regime along $p=0.6$. In \Fig{fig3}(a), the system size scaling of disorder-averaged $\sigma_{xx}$ is shown as a function of $\alpha$, near the DM-AI transition. We fit the usual one-parameter scaling function $\mathcal{F}_1 \sim (\alpha - \alpha_c)L^{1/\nu}$, where $\alpha_c$ is the critical point and $\nu$ is the localization length exponent. We find the best fit for $\alpha_c = -0.294 \pm 0.001$ and $\nu = 2.00 \pm 0.05$ as shown in dashed lines in \Fig{fig3}(a). The scaling collapse for the same is shown in the inset (see SM \cite{sm}). Similarly, we consider another metal-insulator transition where the DM phase eventually enters the CI phase as $p$ is increased along $\alpha = -0.6$. \Fig{fig3}(b) shows the system size scaling of $\sigma_{xx}$ near the transition. Fitting the scaling function $\mathcal{F}_2$ $\sim$ $(p-p_c)L^{1/\nu}$, we extract the critical point at $p_c=0.868\pm 0.001$ with the same exponent $\nu = 2.00 \pm 0.05$ (see collapse in the inset of \Fig{fig3}(b)). While similar thermal metal-insulator transitions driven by energetic disorder in class D TSCs show criticalities with $\nu \approx 1.4$ \cite{Victor_classDexponent_2008, Wang_PRB_2021}, in another disordered two-dimensional SPT with time-reversal symmetry (class DIII) metal-insulator transitions follow $\nu \approx2.7$ universality \cite{Yamakage_PRB_2013}. Thus, the RBCI system exhibits a distinct universality class of metal-insulator transitions, with an exponent $\nu \approx 2.0$.

We further investigate the AI to CI transition in the positive $\alpha$ regime along $p=0.4$. This insulator-insulator transition shows $\nu \approx 1$ universality class (see SM \cite{sm}) as in a Dirac criticality of the topological transition, demonstrating the marginal irrelevance of disorder in our RBCI system. Interestingly, apart from these metal-insulator and insulator-insulator transitions, the RBCI phase diagram also exhibits tricritical points (see \Fig{fig1} and \Fig{fig2}(a)) where all three different phases (DM, AI, and CI) meet. Understanding the nature of such exotic criticality will be an exciting future prospect.

{\it Origin of metal.---}Having discussed various phases and the critical properties of the associated phase transitions, we next investigate the underlying physics behind the emergence of metal in the RBCI. To answer this, let us first consider the AI phase of the geometrically disordered Chern insulator Hamiltonian in the $\alpha\rightarrow0$ limit at an intermediate $p$ (for instance $p=0.5$ cut in \Fig{fig2}(b)). Now, given any configuration of such a random lattice, introducing a finite $\alpha$ {\it reconnects} all the broken bonds in the system. For positive $\alpha$, low-energy localized electrons in the lattice only experience an energetic change and begin to hybridize via {\it stitched} bonds. Eventually, as $\alpha$ increases, hybridization becomes strong enough to gap out all the bulk electrons; only edge states remain in open boundaries due to the topological nature of the parent Hamiltonian. Thus, the localized AI phase undergoes an insulator-insulator transition and enters the CI phase.

\begin{figure}
    \centering
    \includegraphics[width=1\linewidth]{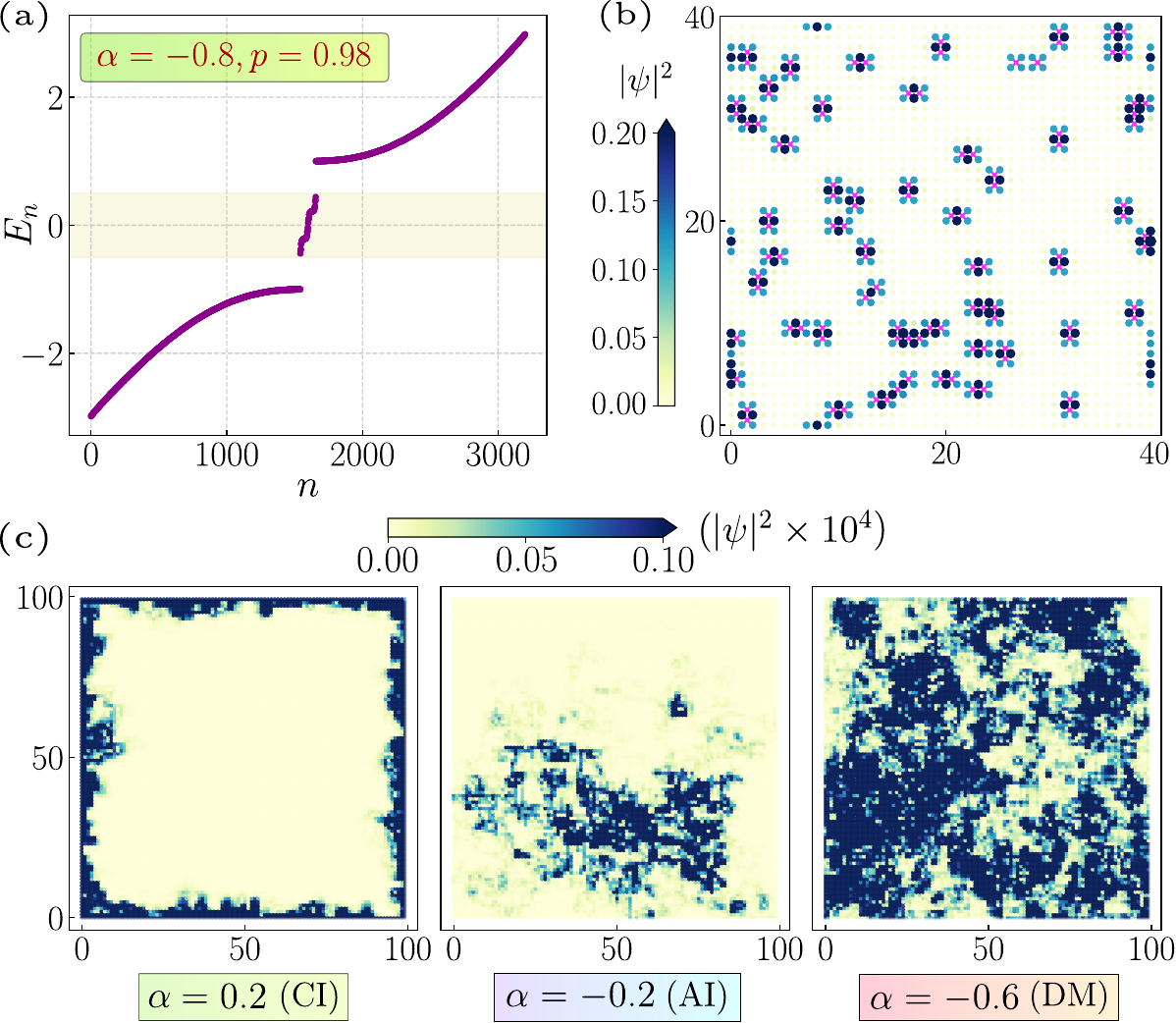}
    \caption{\textbf{Understanding the origin of metal:} (a) Energy spectrum of a typical configuration of the RBCI model with small density of $\pi$-flux defects at $\alpha=-0.8, p=0.98$, shows sub-gap states near zero energy. (b) These states are localized at $\pi$-flux plaquettes in the configuration marked in {\it crosses}. (c) At $p=0.5$, local density of states of near-zero energy modes of a single configuration illustrate all three phases: CI edge state ($\alpha = 0.2)$), localized AI ($\alpha = -0.2$), and extended metallic state ($\alpha=-0.6$).}
    \label{fig4}
\end{figure}

In contrast, for a negative $\alpha$, along with energetic changes, electrons obtain an extra $\pi$ phase when encircle a plaquette with an {\it odd} number of {\it stitched} bonds. This results in randomly distributed $\mathbb{Z}_2$ flux on the lattice. Now, similar to {\it vortices} in TSCs \cite{Kopin_vortex_sc_1991, Volovik_vortex_sc_1999, Read_vortex_sc_2000,  Ivanov_vortex_sc_2001}, $\pi$-flux defects in a topological phase entrap exponentially localized zero-energy modes \cite{Lee_vortexdefect_2007, Qi_vortexinQSH_2008, Ran_vortexinTI_2008, Juricic_vortexprobe_2012} (also see SM \cite{sm}). However, compared to real Majorana fermions in TSCs, these modes are usual complex fermions and carry electronic charge. In \Fig{fig4}(a), we show the spectrum containing such sub-gap modes near zero-energy for a very small density of random $\pi$-flux defects in a typical configuration of RBCI at $\alpha=-0.8$ and $p=0.98$. \Fig{fig4}(b) confirms that these modes are localized near the defects. Now at an intermediate $p$, a large density of such $\pi$-flux plaquettes emerges in the lattice, and their associated zero-modes percolate throughout the systems (see SM \cite{sm}), thus providing conducting pathways for electrons, rendering metallicity in an otherwise localized phase. In \Fig{fig4}(c), such transitions are shown in terms of the local density of near-zero-energy states in a random configuration as $\alpha$ varies from positive to negative values at $p=0.5$. While at $\alpha = 0.2$ only the edge state is present in the open boundary, at $\alpha = -0.2$ flux defects give rise to localized modes in the bulk. With increasing strength of coupling ($\alpha=-0.6$), these modes percolate, giving rise to the metal.

We further investigate the RBCI model on a bond-diluted triangular (see SM \cite{sm}) and find, along with two distinct insulating phases, the diffusive metallic phase arises in the sector where stitching of disconnected clusters generates random $\pi$-flux defects (i.e., negative $\alpha$). This thus demonstrates that the exotic interplay between topology and $\mathbb{Z}_2$ flux is a generic feature of geometrically disordered class D systems.

{\it Outlook.---}Realization of amorphous SPT phases in material \cite{Paul_amorphusbi2se3_2023} and various synthetic settings \cite{LiPRL2021, costa_amorphusti_2019, zhou_amorphusTI_2020, Zhang_amorphustopology_science_2023}, Chern mosaics in twisted-layers solid-state systems \cite{grover_chernmosaic_2022} have shown that the role of geometric disorder is far from ordinary in topological systems. Finding universal phenomena and identifying the role of such disorders in minimal models, even theoretically, can thus provide important insights into both understanding and engineering them. In this work, we have uncovered that generically, class D Chern insulators do host diffusive metals, albeit when disordered in a way that introduces plaquette $\mathbb{Z}_2$ fluxes.

Such diffusive metals, unlike those in superconductors, can carry charge current and even anomalous Hall current. We have further uncovered that the transition between the metal and the Chern insulator is of a different universality class compared to that of the thermal metal, as has been identified in disordered topological superconductors. Thus, our work establishes the profound impact of geometric disorder on the nature of phase transition in disordered SPTs. Natural questions arise regarding the role of dimensionality and anti-unitary symmetries in such random-bond topological models. Moreover, the results will have direct implications on strongly correlated topologically ordered systems where the ground states at finite temperatures experience flux disorder \cite{Perkins_qslfiniteT_2019, Udagawa_qslinfiniteT_2021, Shenoy_piflux_2025}.  We believe our work not only sheds light on the localization-delocalization of fermions in a geometrically disordered class D system, but also lays a platform to explore the effects of random $\mathbb{Z}_2$ gauge fields on disordered topological systems.

{\it Acknowledgment.---} We acknowledge fruitful discussions with Diptarka Das, Ritajit Kundu, Alexander Mirlin, Chandan Dasgupta, Vijay B. Shenoy, and Kedar Damle. S.P. acknowledges funding from IITK Institute Fellowship. N.N. would like to thank DST-INSPIRE fellowship, Grant No. IF- 190078, for funding. A.A. acknowledges support from IITK Initiation Grant (IITK/PHY/2022010). S.B. thanks the National Supercomputing Mission (NSM) for providing computing resources of `PARAM Rudra' at IITB, implemented by C-DAC and supported by the Ministry of Electronics and Information Technology (MeitY) and
Department of Science and Technology (DST), India. Numerical calculations were performed on the workstations \texttt{WIGNER}, \texttt{SYAHI} at IITK.

\bibliography{perco_ref}

\newpage
\setcounter{equation}{0}
\setcounter{table}{0}
\setcounter{figure}{0}
\makeatletter
\renewcommand{\thesection}{S\Roman{section}}
\renewcommand{\theequation}{S\arabic{equation}}
\renewcommand{\thefigure}{S\arabic{figure}}
\renewcommand{\thetable}{S\arabic{table}}
\onecolumngrid

\begin{center}
	\textbf{\large Supplemental Material to ``Diffusive metal in a percolating Chern insulator"}\\
    \vspace{0.4cm}
    {Subrata Pachhal$^1$, Naba P. Nayak$^{2, 3}$, Soumya Bera$^3$, Adhip Agarwala$^1$}
    \\
    \vspace{0.2cm}
    \small{\it $^1$Department of Physics, Indian Institute of Technology Kanpur, Kalyanpur, UP 208016, India}\\
    \small{\it $^2$Institut Langevin, ESPCI Paris, Université PSL, CNRS, 75005 Paris, France}\\
    \small{\it $^3$Department of Physics, Indian Institute of Technology Bombay, Powai, Mumbai 400 076, India}

\end{center}

\vspace{\columnsep}

\twocolumngrid

\section{Virtual crystal approximation of RBCI} \label{sec_mft}
The random bond-disordered Chern insulator (RBCI) system is given by the following two-orbital (namely $A$ and $B$) model Hamiltonian,
\begin{equation}
	H = \sum_{i, \vec{\eta}}\Big( \Psi^\dagger_{i} t_{i, \vec{\eta}} T_{\vec{\eta}}\Psi_{i+\vec{\eta}} + \text{h.c.}  \Big)+ \sum_i \Psi^\dagger_i \sigma_z \Psi_i,
	\label{seq_rham}
\end{equation}
where spinless free fermions $\Psi_{i}^{\dagger} = (c_{iA}^{\dagger}~ c_{iB}^{\dagger})$ hops on a square lattice with unit vector $\vec{\eta} =( \hat{x}$, $\hat{y})$. Associated hopping matrices are $T_{\hat{x}}=-\half \left(\sigma_{z}+i\sigma_{x}\right)$ and $T_{\hat{y}}=-\half\left(\sigma_{z}+i\sigma_{y}\right)$ with $\sigma_{x/y/z}$ representing the Pauli matrices. Bond strengths $t_{i, \vec{\eta}}$ follow the bimodal distribution: $\mathcal{P} (t_{i, \vec{\eta}}) =  p\delta(t_{i, \vec{\eta}} - 1) + (1-p)\delta(t_{i, \vec{\eta}} - \alpha )$, such that given any probability $p$, in a lattice configuration, $p$ fraction of randomly chosen bond has strength $1$ and remaining $(1-p)$ fraction has strength $\alpha \in [-1, 1]$. At $p=1$ and/or $\alpha = 1$, the system is translation invariant, giving rise to the momentum space Hamiltonian,
\begin{align}
\mathcal{H}(\vec{k}) = &~ \sin k_x \sigma_x + \sin k_y \sigma_y \nonumber \\ 
& + (1-\cos k_x - \cos k_y)\sigma_z, \label{seq_qwz}
\end{align} 
known as QWZ model \cite{QWZ_spinlessBHZ_2006}. At half-filling, the system is a Chern insulator with a Chern number $\mathcal{C} = 1$.

\begin{figure}
    \centering
    \includegraphics[width=1\linewidth]{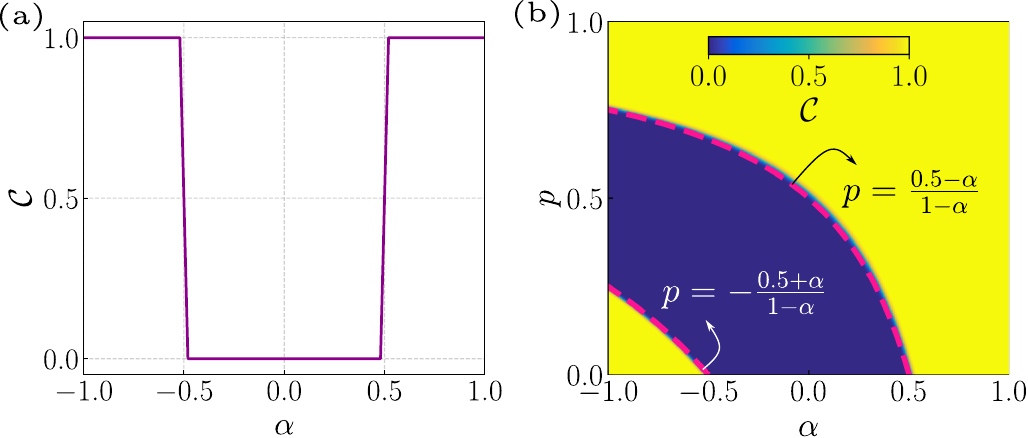}
    \caption{(a) Chern number $\mathcal{C}$ with $\alpha$ for the $p=0$ clean limit Hamiltonian of RBCI (see \eqn{eq_halpha}. (b) Phase diagram of RBCI under virtual crystal approximation (see Hamiltonian in \eqn{eq_hmf}) in terms of $\mathcal{C}$ showing either trivial insulating or Chern insulating phase with $\mathcal{C}=1$.}
    \label{smfig1}
\end{figure}

At {\it zero} probability ($p=0$), translational symmetry gets restored in the system but with re-normalized bond strengths $\alpha$, thus,
\begin{align}
\mathcal{H'}(\vec{k}) = &~ \alpha \sin k_x \sigma_x + \alpha \sin k_y \sigma_y \nonumber \\ 
& + (1- \alpha \cos k_x - \alpha \cos k_y)\sigma_z.\label{eq_halpha}
\end{align}
$\mathcal{H'}(\vec{k})$ has two topologically distinct insulating phases: $\mathcal{C} = 0$ for $|\alpha| < 0.5$ and $\mathcal{C} = 1$ otherwise, separated by Dirac gap closing at $\Gamma \equiv (0, 0)$ point for $\alpha = 0.5$ and $(\pi, \pi)$ point for  $\alpha =-0.5$ (see \Fig{smfig1}(a)).

Now, to capture only the energetic correction to these phases at a given $0<p<1$, we consider the virtual crystal approximation \cite{Soven_CPA_PR1967}. Ignoring any geometric fluctuations, we replace all the hopping with the average strength $\tilde{\alpha} = p + (1-p)\alpha$, and get the following translationally invariant Hamiltonian, 
\begin{align}
\tilde{\mathcal{H}}(\vec{k}) = &~ \tilde{\alpha} \sin k_x \sigma_x + \tilde{\alpha} \sin k_y \sigma_y \nonumber \\ 
& + (1- \tilde{\alpha} \cos k_x - \tilde{\alpha} \cos k_y)\sigma_z.\label{eq_hmf}
\end{align}
Thus, in the $\alpha$-$p$ phase diagram, the critical lines separating topological and trivial phases under virtual crystal approximation are given by, $\tilde{\alpha} = \pm 0.5$, thus $p = (\pm 0.5 - \alpha)/(1-\alpha)$.  Numerically calculated Chern number phase diagram shown in \Fig{smfig1}(b) confirms the same.

\begin{figure*}
    \centering
    \includegraphics[width=0.8\linewidth]{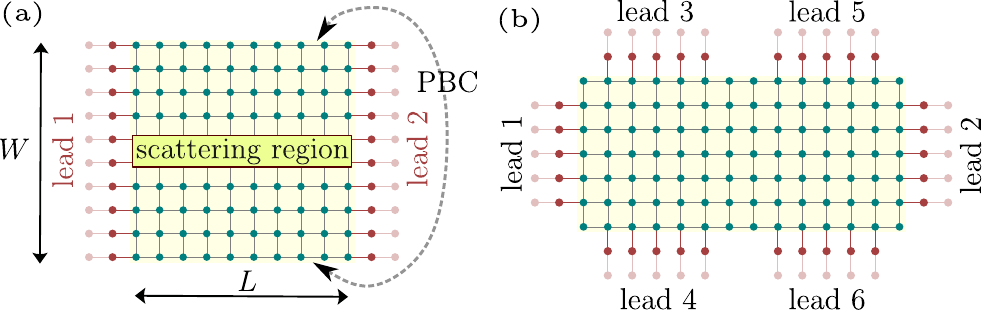}
    \caption{(a) Two-terminal setup where a scattering region (here a square lattice) of length $L$ and width $W$. Two leads, composed of a series of one-dimensional wires, are connected to the left and right edges. The transverse edges are periodic (PBC).  (b) Six-terminal geometry where we always choose $L = 2W$, such that all the $6$ leads are uniformly placed.}
    \label{smfig2}
\end{figure*}

\section{Conductivity calculation using \texttt{KWANT}}
To calculate conductivities, both longitudinal ($\sigma_{xx}$) and transverse ($\sigma_{xy}$), we employ  quantum transport package \texttt{KWANT} \cite{Groth_Kwant_2014}. For $\sigma_{xx}$, a two-terminal setup is used with the periodic boundary condition (PBC) in the transverse direction, as shown in \Fig{smfig2}(a). Two leads connected to the left and right edges are essentially a bunch of semi-infinite one-dimensional metallic channels with electronic states available at the Fermi energy $E_F = 0$. As the model of our study has two orbital degrees of freedom, at each site of the left and right edges, two such disconnected channels are attached. In this setup, calculating the scattering matrix yields the transmission $T_{12}$ between lead 1 and lead 2. Using the Landauer-Buttiker formula, the conductance is $G_{12} = (e^2/h)T_{12}$ \cite{Datta_Book_1997}. Thus, given the length and width of the scattering region are $L$ and $W$ respectively, the longitudinal conductivity is, 
\begin{equation}
    \sigma_{xx} = \frac{L}{W} G_{12} = \frac{L}{W}T_{12}\frac{e^2}{h}.
\end{equation}

For the $\sigma_{xy}$ calculation, we use a six-terminal device as shown in \Fig{smfig2}(b) and consider $L = 2W$ with uniformly placed leads to reduce the geometric effects. Calculating the transmission $T_{mn}$ between lead $m$ and $n$ using \texttt{KWANT} (note $T_{mm} = 0$), we get the current at different leads from the following relation,
\begin{equation}
I_m =  \frac{e^2}{h} \sum_{n\neq m}\big(T_{nm}V_m - T_{mn}V_n\big),\label{eq_sixterminal} 
\end{equation}
where $m$ and $n$ run from 1 to 6 and $V_{m/n}$ is voltage at lead $m/n$. While leads 1 and 2 are used as the current input, leads 3, 4, 5, and 6 are used as voltage probes. To force these conditions, we fix $V_1 = 1, V2 = 0$ and $I_{3/4/5/6} = 0$ such that $I_1 = -I_2$ and $V_{3/4/5/6}$ are the unknown voltages. Solving \eqn{eq_sixterminal}, we get these unknown quantities and calculate longitudinal and transverse resistances as follows, 
\begin{align}
R_{xx} = \half \left( \frac{V_3 - V_5}{I_1} + \frac{V_4 - V_6}{I_1} \right), \\
R_{xy} = \half \left( \frac{V_4 - V_3}{I_1} + \frac{V_6 - V_5}{I_1} \right).
\end{align}
Using these we form the resistance tensor with $R_{yy} = R_{xx}$ and $R_{yx} = - R_{xy}$. The averaging further reduces the effects of anisotropic geometry. Finally, inverting the resistance tensor, we get the Hall conductivity $\sigma_{xy}$ in units of $\frac{e^2}{h}$.

\section{Density of states}
The DOS provides information about the number of available single-particle states per unit energy. 
For a Hamiltonian $H$ with eigenvalues $\{E_n\}$, the density of states (DOS) is defined as,
\begin{equation}
\rho(E) = \frac{1}{N} \sum_{n} \delta(E - E_n),
\end{equation}
where $N$ is the total number of lattice sites. The DOS can be efficiently evaluated using the Kernel Polynomial Method (KPM)~\cite{KPM_RMP_2006}. In this approach, after rescaling the Hamiltonian spectrum to the interval $[-1,1]$,  the DOS is expanded in terms of the Chebyshev polynomials as follows, 
\begin{equation}
\rho(E) \approx \frac{1}{\pi \sqrt{1-E^2}}
\left[
g_0 \mu_0 + 2 \sum_{m=1}^{N_m-1} g_m \mu_m T_m(E)
\right],
\end{equation}
where $T_m(E)$ are the Chebyshev polynomials of the first kind, $\mu_m$ are the Chebyshev moments, $g_m$ are kernel coefficients, and $m$ is the truncation order of the expansion. The DOS at a specific energy, in particular at zero energy $\rho(E=0)$, can be systematically improved by increasing the number of moments $N_m$, which reduces finite resolution effects inherent to the polynomial expansion. In this work, we employ the KPM with increasing expansion order to obtain converged estimates of the DOS.

\begin{figure}
    \centering
    \includegraphics[width=1\linewidth]{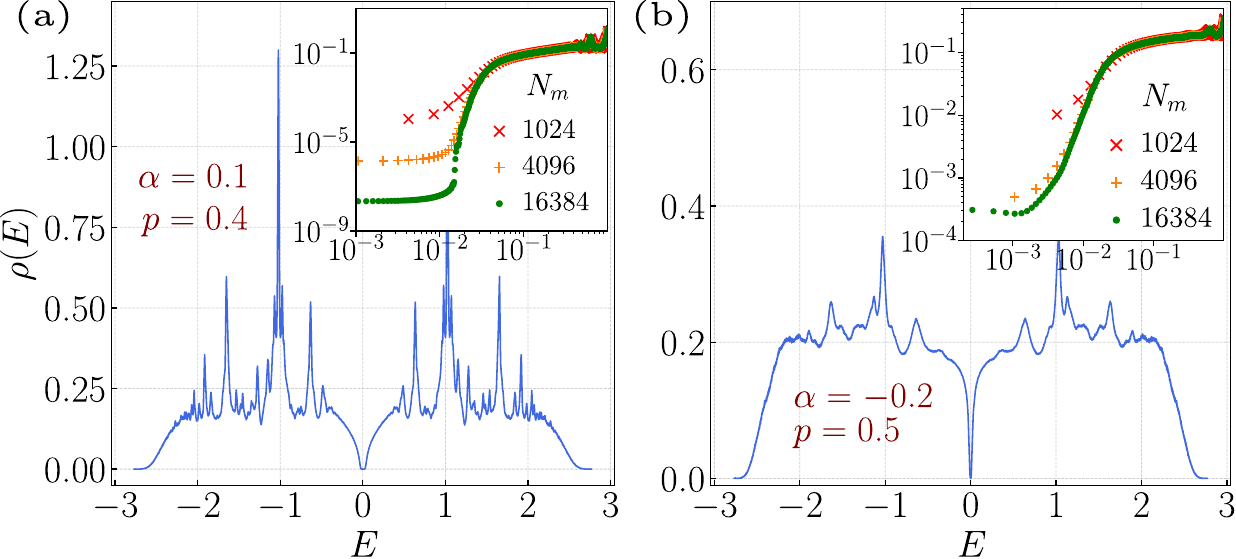}
    \caption{DOS in the localized phase with parameters (a) $p=0.4, \alpha=0.1$ and (b) $p=0.5, \alpha=-0.2$ considering $L=2048$ and $N_m = 16384$. In the insets of both (a) and (b), the corresponding DOS are plotted for various moments $N_m$ of the expansion.}
    \label{fig:dos}
\end{figure}

The DOS diverges in the diffusive metal phase, as shown for a point in the metallic region of the phase diagram in \Fig{fig2}(d) in the main text. In the Anderson insulating region, DOS for $p=0.4, \alpha=0.1$ is shown in \Fig{fig:dos}(a). The result shows a clear bulk gap for the largest system size studied. The inset shows the low-energy density of states to the scale of $10^{-4}$ obtained using the KPM. The gapped system is marked by the saturation of the zero energy density of states $\rho(0)$ to a value capped by $~1/N_m.$ The gap is reduced significantly for the point $p=0.5, \alpha=-0.2$ as shown in \Fig{fig:dos}(b).

\section{Inverse participation ratio}
To capture the nature of the eigenstates in the localized Anderson insulator and the metallic phase, we numerically evaluate the inverse participation ratio (IPR) by employing exact diagonalization. Generalized IPR or $q^{\text{th}}$-moment of IPR of a normalized wavefunction $\psi_{if}$ is defined as,
\begin{equation}
I_q = \sum_{i,f} |\psi_{if}|^{2q},
\end{equation}
where $i$ indicates the unit cells of the lattice and $f$ denotes orbitals at each unit cell.

\subsection{IPR in the Anderson insulator phase} We evaluate the usual IPR $I_{(q=2)}$ for a single configuration of the RBCI model on a $40 \times 40$ square lattice. In \Fig{smfig4}(a) and \Fig{smfig4}(b), we show $I_{(q=2)}$ of the full spectrum, for two points in the Anderson insulator phase (parameter values are the same as in \Fig{fig:dos}(a) and \Fig{fig:dos}(b) respectively). The higher values of IPR, $I_{(q=2)} \sim (0.005 - 0.01)$, in both \Fig{smfig4}(a) and \Fig{smfig4}(b) suggest most of the eigenstates are localized within 100-200 sites out of 1600 sites. This thus demonstrates Anderson localization, although DOS shows a small gap.

\subsection{Multifractal states of diffusive metal}
\begin{figure}
    \centering
    \includegraphics[width=1\linewidth]{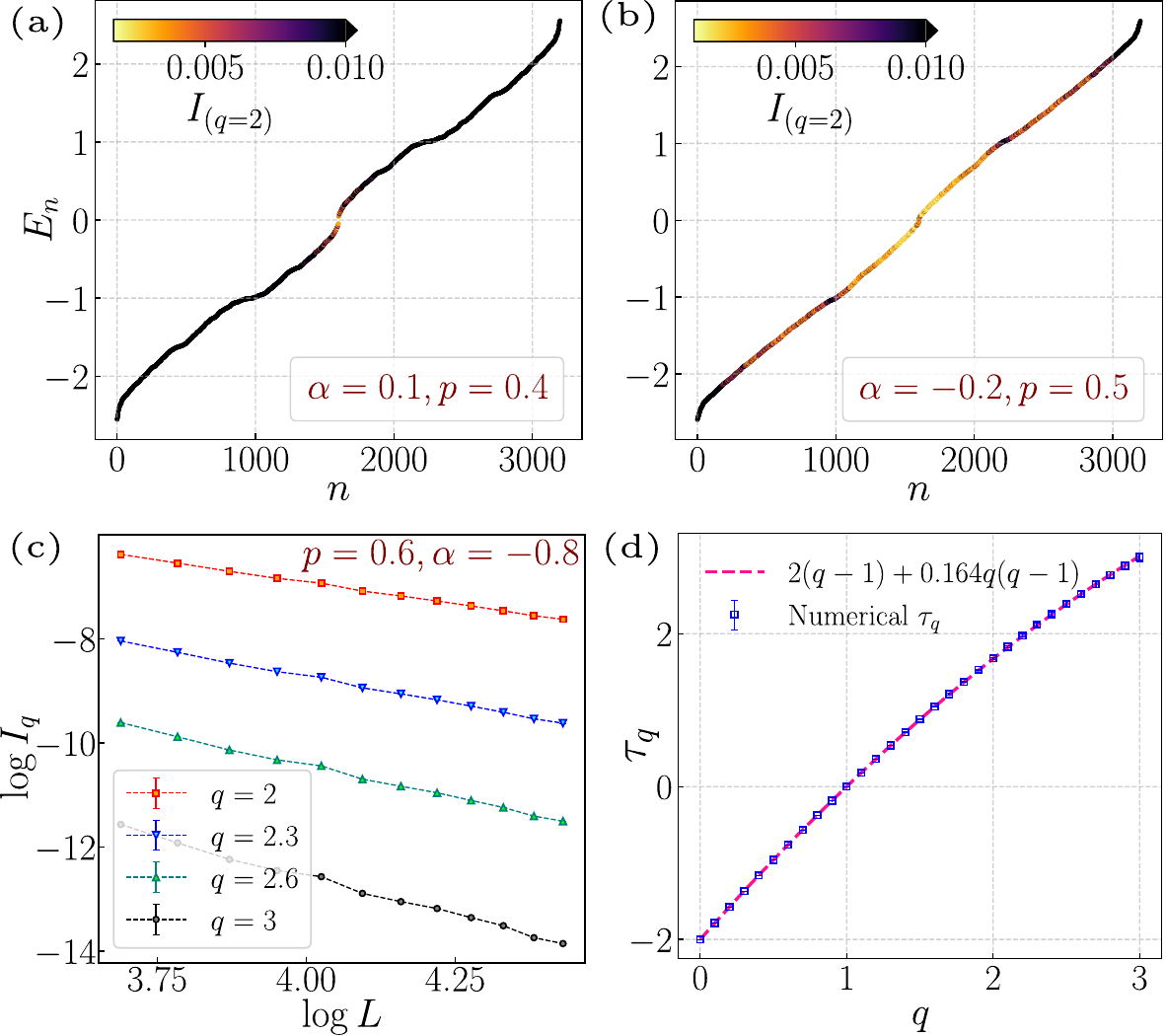  }
    \caption{IPR of the full spectrum in a single configuration of the RBCI with parameters (a) $\alpha = 0.1, p= 0.4$ and (b) $\alpha = -0.2, p= 0.5$. For both (a) and (b) linear system size is $L=40$. (c) At parameter value $p=0.6, \alpha=-0.8$, system size scaling of configuration-averaged $I_q$ of a near-zero energy eigenstate for various $q$. The average is taken over 200 configurations. (d) The scaling exponent $\tau_q$ (see \eqn{seq_tauq}) as a function of $q$ exhibits multifractal behavior.}
    \label{smfig4}
\end{figure}

To see the nature of the metallic eigenstate, we calculate the generalized IPR $I_{q}$. In \Fig{smfig4}(c), we show the scaling of configuration-averaged $I_q$ of a near-zero energy eigenstate for $p=0.6, \alpha =-0.8$ for various $q$. They scale with linear system size $L$ as 
\begin{align}
    I_q \sim L^{-\tau_q}. \label{seq_tauq}
\end{align}
Employing numerical fitting, the exponent $\tau_q$ is evaluated as a function of $q$, shown in \Fig{smfig4}(d). The result shows a scaling behavior,
\begin{align}
    \tau_q = 2(q-1) + \gamma q (1-q),
\end{align}
with multifractality exponent $\gamma \approx 0.164$. This demonstrates the multifractal nature of the metallic states in the DM phase.

\section{Fitting procedure}
In this section, we outline the fitting procedure employed in the main text to estimate the critical parameters associated with quantum phase transitions. In general, the observables are expanded in terms of both relevant and irrelevant scaling variables up to a certain order, and the resulting scaling function is then fitted to the numerical data for finite-size scaling analysis. To perform the fit, the expansion can be truncated at a chosen order in these variables. In our study, we truncate the irrelevant-variable expansion at second order to avoid overfitting while capturing the leading corrections. The expansion of the order parameter (here, disorder disorder-averaged conductivity $\overline{\sigma}$) is given as follows,

\begin{equation}
\overline{\sigma} = f_0(x) + b_0 L^{-y} f_1(x) + c_0 L^{-2y} f_2(x) 
\label{eq:g_expand}
\end{equation} 
where, 
$
f_j(x) = \sum\limits_{n=0}^{N_\text{R}} a_{j n} x^n,
$
$x = (M-M_\text{c})/M_\text{c} \cdot  L^{1/\nu}$, $M$ being the parameter that drives the transition ($M_c$ the critical value), $\nu$ and $y$ are the leading relevant and irrelevant exponents respectively and $(b_0,c_0,a_{jn})$ are expansion coefficients. $N_R$ is the order of relevant expansion that can be varied to best fit the numerical data and prevent over-fitting.

\begin{table}
\renewcommand{\arraystretch}{1.4}
\centering
\begin{tabular}{p{38mm} ccccccc}
\toprule
\toprule
 & \multicolumn{7}{c}{$L$} \\
\cmidrule(lr){2-8}
Data  & 64 & 96 & 128 & 192 & 256 & 384 & 512 \\
\midrule
($\sigma,\alpha$) (\Fig{fig3}(a)) 
 & 1.5 & 1.5 & 0.5 & 0.5 & 0.2 & 0.2 & -- \\

($\sigma,p$) (\Fig{fig3}(b)) 
 & 0.8 & -- & 0.32 & -- & 0.16 & -- & 0.1 \\

($\sigma,\alpha$) (\Fig{smfig5}(a)) 
 & 0.8 & 1.5 & 0.32 & 0.5 & 0.16 & 0.1 & -- \\
\bottomrule
\bottomrule
\end{tabular}

\caption{Disorder configurations (in units of $\times 10^{5}$) used for different observables and varying parameters $\alpha$ and $p$. The lattice width is fixed to $W=2L$ for all realizations.}
\label{tab:dis_config}
\end{table}

The fitting of the function to the data is performed using a non-linear least-squares approach. We have examined the robustness of the extracted critical exponents under different choices of $N_R$ by minimizing the chi-square of the fit, defined as
\begin{equation}
    \chi^2 = \sum_{i=1}^{N_D} \frac{\left(F_i - \Gamma_i\right)^2}{\Gamma_i},\label{chi_square}
\end{equation}
where $F_i$ denotes the value of the scaling function evaluated at the $i^{\text{th}}$ data point, $\Gamma_i$ represents the corresponding numerical measurement, and $N_D$ is the total number of data points. Error bars are obtained through bootstrap resampling of the dataset. We also note that the irrelevant exponent $y$ is large for metal–insulator critical points and can thus be neglected, whereas it remains finite for insulator–insulator transitions.

The table below provides details on the number of configurations used to determine the critical properties for various observables and parameters (see Table \ref{tab:dis_config}).

\section{Insulator-insulator phase transition}
\begin{figure}
    \centering
    \includegraphics[width=1\columnwidth]{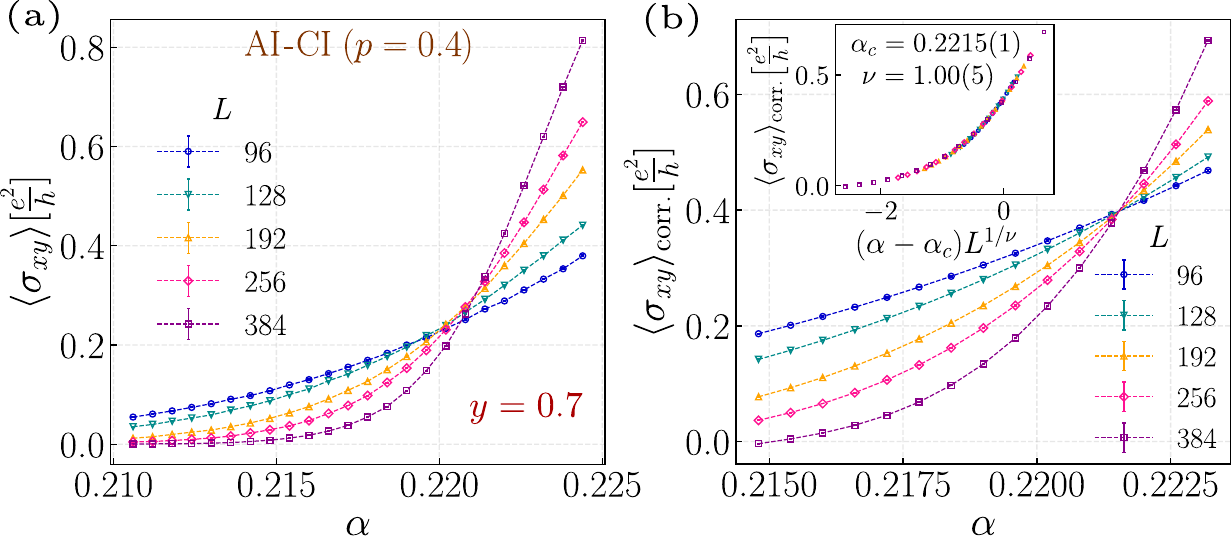} 
    \caption{(a) Scaling of raw $\sigma_{xx}$ data with $\alpha$ near the AI-CI transition at $p=0.4$ shows a shift in the crossing point, indicating strong finite size correction with exponent $y=0.7$. (b) The corrected $\langle \sigma_{xx} \rangle_{\text{corr.}}$ and their collapse (inset) show transition at $\alpha_c \approx 0.2215$ with $\nu \approx 1$. For all the data points, the disorder average is calculated over $4\times 10^4$ configurations, and error bars are within the width of the markers.}
    \label{smfig5}
\end{figure}

In this section, we investigate the Anderson insulator (AI) to Chern insulator (CI) transition in the positive $\alpha$ regime at $p=0.4$. The raw $\sigma_{xx}$ data shown in \Fig{smfig5}(a) exhibit a shift in the crossing point with consecutive system sizes, suggesting a strong finite size correction \cite{Slevin_ycorrection_2009} with irrelevant exponent $y=0.7$ (see \eqn{eq:g_expand}). Using corrected data $\langle \sigma_{xx} \rangle_{\text{corr.}}$ by subtracting the irrelevant part, we find that the AI-CI transition for $p=0.4$ occurs at $\alpha_c = 0.2215 \pm 0.0001$ and has an exponent $\nu = 1.00 \pm 0.05$ (see \Fig{smfig5}(b) and its inset). Thus, the insulator-insulator transition in RBCI is in the same universality class as in a clean Dirac criticality ($\nu = 1$), demonstrating the marginal irrelevance of disorder.

\section{Flux defect in a Chern insulator}

\begin{figure*}
    \centering
    \includegraphics[width=1\linewidth]{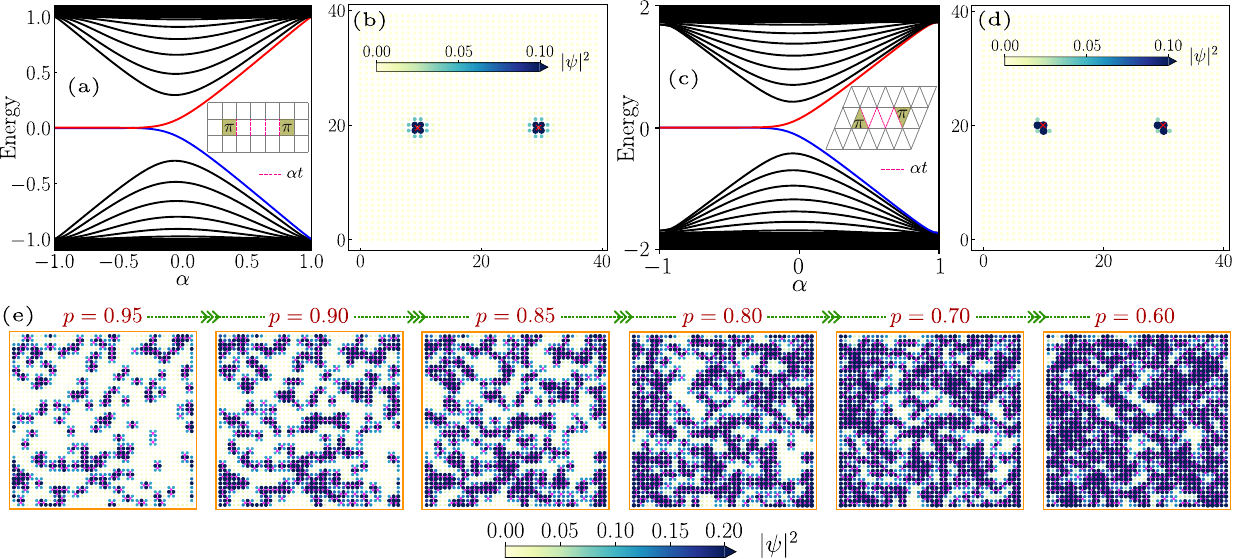} 
    \caption{(a) Sub-gap modes due to $\pi$-flux defect in a Chern insulator. In a $40\times 40$ square lattice, two flux defects are created at a 20 unit cell distance, using the $\alpha$ parameter as schematically shown in the inset. (b) The local density of the zero-energy modes for $\alpha=-0.8$ indicates that they reside on the defect plaquettes, as shown by the {\it cross} marks. (c) and (d) are the same as (a) and (b), respectively, but for a triangular lattice Chern insulator. (e) Local density of zero-modes in random configurations of the RBCI ($L=40$) for different $p$ values at $\alpha = -0.8$. With decreasing $p$, the density of random $\pi$-flux plaquettes ({\it cross} mark) increases, and their corresponding zero-modes spread throughout the system.}
    \label{smfig6}
\end{figure*}

While in a topological superconductor, vortices bind Majorana zero modes \cite{Kopin_vortex_sc_1991, Volovik_vortex_sc_1999, Read_vortex_sc_2000, Ivanov_vortex_sc_2001}, in two-dimensional topological systems such as quantum spin Hall insulators, a pair of zero-energy states emerges in the presence of a $\pi$-flux defect, which are exponentially localized \cite{Lee_vortexdefect_2007, Ran_vortexinTI_2008, Qi_vortexinQSH_2008, Juricic_vortexprobe_2012}. Similarly, in a class D Chern insulator, these flux defects create zero-energy modes in the system. Considering the clean Chern insulator as described in \eqn{seq_qwz}, we insert flux defects on the plaquettes by flipping the sign of the bond separating the plaquettes. Using a string of bond flipping, we can create these defects at any plaquettes in the configuration as shown in the inset of \Fig{smfig6}(a) with $\alpha = -1$. Now varying $\alpha$ from -1 to 1, we show the spectrum in \Fig{smfig6}(a) contains two sub-gap states due to the defects, and they remain near zero-energy for $\alpha<0$ and gap out for $\alpha \geq 0$. These modes are attached to the defect plaquettes, as shown in \Fig{smfig6}(b) for $\alpha = -0.8$. Note that for flux defects, zero modes only arise when the system is in a topologically non-trivial phase, in contrast to the case where potential defects can create sub-gap states even in a trivial insulator \cite{Wimmer_vortex_prl_2010, Pachhal_topowire_PRB_2025}.

We also consider a triangular lattice Chern insulator to demonstrate the universality of the physics of the $\mathbb{Z}_2$ flux defects. In momentum space, the triangular lattice Hamiltonian is given by,
\begin{align}
	\mathcal{H}_{\text{Tri}}(\vec{k}) = &~   \bigg(1-\cos k_x - \cos \frac{k_x}{2} \cos \frac{\sqrt{3}k_y}{2} \bigg)\sigma_z  \nonumber \\ & - \bigg(\sin k_x + \sin \frac{k_x}{2} \cos \frac{\sqrt{3}k_y}{2}\bigg) \sigma_x \nonumber \\  & - \bigg( \sqrt{3}\cos \frac{k_x}{2} \sin \frac{\sqrt{3}k_y}{2} \bigg) \sigma_y.\label{eq_tribhz}
\end{align}
Using a similar protocol as described above, we create two $\pi$-flux defects in the system using the $\alpha$ parameter. The spectrum with $\alpha$ is shown in \Fig{smfig6}(c) showing zero-modes for $\alpha < 0$. \Fig{smfig6}(d) shows the local density of these modes localized at the defect plaquettes.

In the RBCI system, for negative $\alpha$, given a $p$, a density of randomly distributed $\mathbb{Z}_2$ flux defects appears in a configuration, and the energy spectrum exhibits associated defect-induced modes near zero energy. Considering $\alpha = -0.8$, we show the evolution of such a configuration of these modes with $p$ in \Fig{smfig6}(e) starting from the CI phase at $p=0.95$. As the $p$ decreases, the density of random $\pi$-flux increases, and their associated zero-modes percolate throughout the system, giving rise to conducting channels. Thus, the configuration goes through a metal-insulator transition and becomes metallic.

\section{Triangular lattice RBCI}
\begin{figure}
    \centering
    \includegraphics[width=1\columnwidth]{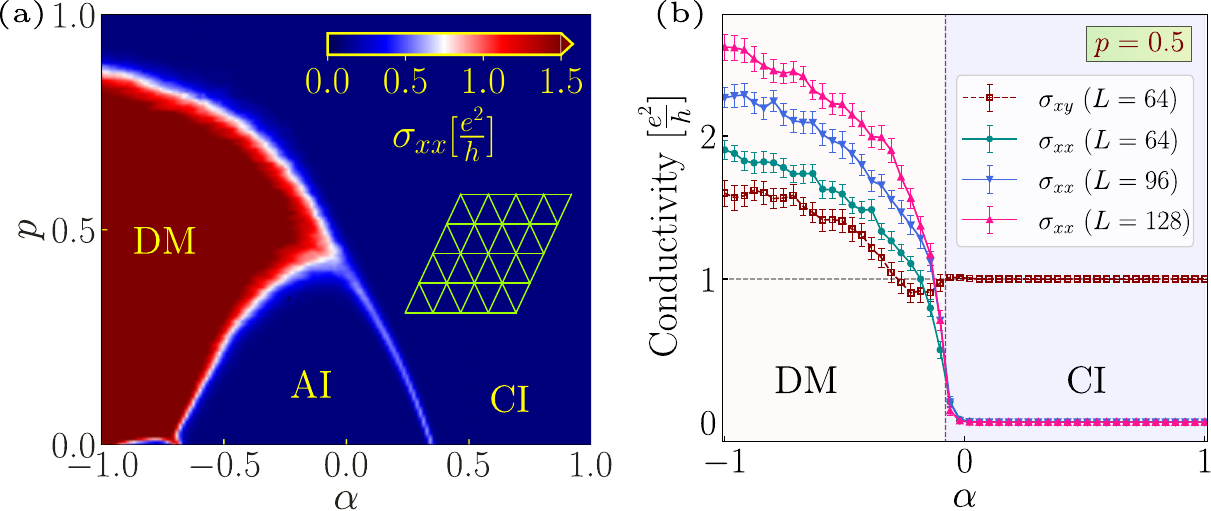} 
    \caption{(a) The phase diagram of the RBCI model in a $40 \times 40$ triangular lattice in terms of $\sigma_{xx}$ (averaged over $200$ configurations). (b) Distinct phases along the cut $p=0.5$, characterized by $\Delta E, \sigma_{xx}$ and $\sigma_{xy}$, show a similar DM phase as in the square lattice case. All data in (d) are averaged over 200 disorder realizations.}
    \label{smfig7}
\end{figure}

Let us consider a percolating Chern insulator model with random bonds on a triangular lattice. The clean model at $p=1$ is given by \eqn{eq_tribhz}. Similar to the square lattice case, the phase diagram of the triangular lattice RBCI consists of a diffusive metal (DM) phase, as shown in \Fig{smfig7}(a), along with Anderson insulator (AI) and Chern insulator (CI) phases. \Fig{smfig7}(b) shows the diffusive nature in $\sigma_{xx}$ and a non-vanishing $\sigma_{xy}$ in the DM phase. Importantly, the DM phase arises in the negative $\alpha$ sector where stitching of disconnected clusters generates random $\pi$-flux plaquettes in the system. It is pertinent to note that, unlike the square lattice RBCI model in the triangular lattice, there is no $p \leftrightarrow (1-p)$ duality along $\alpha = -1$. While in a square lattice, for $\alpha=-1$, both $p=0$ and $p=1$ correspond to zero-flux configurations for all the plaquettes. In a triangular lattice, they correspond to zero and $\pi$-flux, respectively.

\end{document}